# Experimental Simulation of Dust Impacts at Starflight Velocities

Andrew J. Higgins[1]

[1] Department of Mechanical Engineering, McGill University, Montréal, Québec, Canada

## Abstract

The problem of simulating the interaction of a spacecraft travelling at velocities necessary for starflight with the interplanetary and interstellar medium is considered. Interaction of protons, atoms, and ions at kinetic energies relative to the spacecraft (MeV per nucleon) is essentially a problem of sputtering, for which a wealth of experimental data exists at the velocities of interest. More problematic is the impact of dust grains, macroscopic objects on the order of 10 nm ($10^{-21}$ kg) to 1 μm ($10^{-15}$ kg) and possibly larger, the effects of which are difficult to calculate from first principles, and thus experiments are needed. The maximum velocity of dust grains that can be achieved at present in the laboratory using electrostatic methods is approximately 100 km/s, two orders of magnitude below starflight velocities. The attainment of greater velocities has been previously considered in connection with the concept of impact fusion and was concluded to be technologically very challenging. The reasons for this are explained in terms of field emission, which limits the charge-to-mass ratio on the macroscopic particle being accelerated as well as the voltage potential gradient of the accelerating electrostatic field, resulting in the accelerator needing to be hundreds to thousands of kilometers long for μm-sized grains. Use of circular accelerators (e.g., cyclotrons and synchrotrons) is not practical due to limitations on magnetic field strength making the accelerator thousands of kilometers in size for μm-sized grains. Electromagnetic launchers (railguns, coilguns, etc.) have not been able to produce velocities greater than conventional gas guns (< 10 km/s). The nearest feasible technologies (tandem accelerators, macromolecular accelerators, etc.) to reach the regime of projectile mass and velocity of interest are reviewed. Pulsed laser facilities are found to be the only facilities able to accelerate condensed phase matter to velocities approaching 1000 km/s but are unlikely to be able to reach greater speeds. They also cannot create well-quantified, free-flying projectiles. Instead, it is proposed to use pulse laser facilities to simulate the plasma "fireball" that results from such impacts, rather than try to reproduce the impacts themselves. Pulsed laser facilities exist that can provide the energy and power densities to recreate the consequences of an impact (if not an impact itself) in the lab. By performing time-resolved measurements of the effects experienced by material samples at scaled distances from the laser-driven event, the damage to representative spacecraft structures can be accurately assessed.

**Nomenclature**
$B$      magnetic field (T)
$c$      speed of light
$d$      diameter (m)
$m$      particle or projectile mass (kg)
$q$      charge (C)
$r$      radius (m)
$v$      velocity (m/s)
$V$      voltage potential (V)
$\delta$      ratio of wall thickness to shell radius
$\varepsilon_0$      permittivity of free space (C$^2$/N m$^2$)
$\mu_0$      magnetic permeability of free space (N/A$^2$)
$\rho_p$      particle or projectile density
$\sigma_c$      surface charge (C/m$^2$)

σ$_t$     tensile stress (N/m$^2$)
Subscripts
FE     field emission
t       tensile
ult     ultimate strength

## 1.     Introduction

Spacecraft traveling at velocities necessary for interstellar travel will encounter interplanetary and interstellar media with potentially devastating consequences.[1-7] The interplanetary and interstellar media encountered consist of both gas (atoms, ions, molecules) and macroscopic particles (dust grains). The response of spacecraft surfaces to the impact of individual atoms, ions, and molecules is essentially a problem of *sputtering*, a phenomenon that has been extensively studied.[8] Sputtering is the ejection of material from a surface via the impact of atoms or ions due to mechanisms distinct from bulk thermalization of the impact energy, essentially, surface-layer atoms being knocked-off the target material. Sputtering has numerous technological applications, such as thin layer deposition, secondary ion mass spectrometry, and ion implantation, and thus has been widely investigated. Sputtering yield, defined as the number of atoms ejected per incident ion, is usually in the range of $10^{-3}$ to $10^1$, generally increasing as the energy of the impacting ion increases. As the ion energies approach 1 MeV[1], wherein mechanism of sputtering becomes dominated by interaction with the electron cloud of the target—termed electronic sputtering—the values of yield typically remain less than $10^2$ for metallic targets, although the yield can be orders of magnitude greater (as great as $10^5$) for insulating targets.[9] Using a sputtering yield value of $10^1$ and assuming the Local Interstellar Cloud has an ion density of 1 ion/cm$^3$, then a light year of travel would result in $10^{20}$ atoms/cm$^2$ being sputtered off forward-facing spacecraft surfaces, which would only correspond to a thickness on the order of a micron of material. While damage to sensitive components (e.g., electronics or optical surfaces) is a valid concern, overall structural degradation of spacecraft structures on the scale of mm thickness due to interaction with the interstellar medium does not appear an insurmountable problem.

More worrisome is the impact of grains, macroscopic objects that can exceed 1 μm in size. At speeds of 0.1 *c*, a 1-μm-diameter dust grain ($10^{-15}$ kg) has approximately 0.5 J of kinetic energy relative to a target, which is equivalent to the energy released by detonation of approximately 0.1 mg of explosives, not a trivial amount. As argued by Schneider [6], if even a single 100 μm grain ($10^{-9}$ kg) were encountered over the duration of the trip, the total kinetic energy of the impactor would be equivalent to 100 g of explosives, and if even a fraction of this energy couples to the spacecraft structure being impacted, such as a light sail, the vehicle may be compromised. For a classic light sail mission [10-12] wherein the acceleration phase occurs over distances measured in light years (LY), the number of impacts of dust grains expected in interstellar space is essentially the volume swept out by the sail multiplied by the number density of grains. The mass density of dust in the local interstellar medium is $3 \times 10^{-23}$ kg/m$^3$.[13] If an average size of 1 μm and density of 2 g/cm$^3$ is assumed, the number density is $3 \times 10^{-8}$ grains/m$^3$, yielding 30,000 impacts/cm$^2$ over a light year of travel, giving a mean spacing between impacts of 60 μm. This estimate is consistent with those of Landis [3] and Early and London [4]. Clearly, if the region damaged by the impact exceeds the grain size by a factor of more than 10, there will be no sail left after a LY of travel. Recently, Lubin [14] has suggested performing the acceleration phase of a laser-driven sail over a much shorter distance, less than an AU, to avoid the problem of collimating the laser over long distances by exploiting highly reflective sail material. This concept has now become the basis for Breakthrough Starshot. The acceleration in this case would occur in the near Earth-orbit region of the solar system, where the mass density of dust is greater ($10^{-20}$ kg/m$^3$).[13] The number of impacts expected would be 140 impacts/cm$^2$ per AU if again 1 μm dust grains are assumed, giving a mean spacing between impact points of approximately 1 mm. Interestingly, this is on the order of the size of the proposed *chipsat* (or *starchip*) that would be accelerated by the light sail, meaning the chipsat may be able to avoid impacts during the acceleration phase and then be oriented edge-on for interstellar coasting. The sail (dimensions measured in m) would not be able to avoid impacts during the acceleration process and is thus a significant concern.

---

[1]As discussed later in this paper, energies of MeV per nucleon correspond to velocities of 5% of light speed.

The problem of macroscopic objects impacting at velocities in the range of 1000 to 30,000 km/s, i.e., up to 10% light speed, has received only very preliminary theoretical consideration and has never been studied experimentally, for reasons that will be explained in this paper. The closest application that has been considered for macroscopic impactors at these velocities is that of impact fusion, a concept that originated with Harrison [15] and Winterberg [16] and generated some enthusiasm in the 1960s and 1970s. Impact fusion would involve accelerating pellets on the order of a gram to 100 to 1000 km/s, still an order of magnitude below starflight velocities. The impact fusion concept has not been attempted experimentally.

Experience from more prosaic hypervelocity impacts of micrometeoroids and orbital debris (MMOD), while likely of little direct relevance to impacts of dust at starflight velocity, well demonstrates the complex physics that can occur under the extreme conditions generated by impact. The typical double-wall shielding used in orbit today (i.e., Whipple bumper) can result in successive regions of "penetration" or "no penetration" as the velocity of a fixed-sized impactor is increased from 2 to 10 km/s due to tradeoffs in impact-generated pressures and temperatures and material strength. The physics of impacts exceeding just 10 km/s are poorly understood—a regime wherein the impactor and target material are expected to vaporize upon release of the impact-generated shock—due to a lack of constitutive data (equation of state, material strength, etc.) and is a direct consequence of our inability to simulate such impact speeds in the laboratory. That different—potentially novel—regimes of impact physics will be encountered as the velocity increases from 10 km/s to 30,000 km/s should be taken as a given.

One such example of possible new phenomena, originally suggested by Winterberg [17], that might occur as impact velocities begin to exceed 100 km/s (generating pressures exceeding 10 TPa) is the formation of transient, metastable inner shell molecules that, upon decay, would emit copious soft X-rays. This phenomenon was suggested by Bae to be the mechanism responsible for anomalous signals observed at Brookhaven National Laboratory in 1994 examining impact of electrostatically accelerated bio and water particles at ~100 km/s.[18] Further, Bae suggests that since these excited states exist in particles smaller than the wavelength of light emitted, they would decay via the mechanism of Dicke superradiance, which would occur orders of magnitude faster than the usual optical decay. This phenomenon—albeit highly speculative—would significantly alter how the impact-generated plasma would interact with the rest of the spacecraft structure assumed in prior calculations of impact damage.

Thus, it is difficult to conceive that an interstellar mission would be undertaken when it is likely that new physics will be encountered as the spacecraft are impacted by dust grains with problematic or impossible-to-calculate consequences for the survivability of the spacecraft. While more grandiose interstellar mission architectures, such as Project Daedalus, have suggested dust shields that would proceed the spacecraft or active sensing and deflection of dust, the question of what will happen if one grain gets past the shield will always arise. Therefore, the ability to simulate dust grain impacts in the laboratory would be of great interest and may well be crucial to the development of interstellar capability. This paper will review the potential for existing or near-future experimental capabilities to address this problem in the laboratory.

## 2.     Accelerator Technologies

Detailed overviews of accelerator technologies capable of accelerating macroscopic objects to the $10^2$–$10^3$ km/s regime can be found in Manzon [19] and in a more layman-oriented article by Kreisler [20]. A thorough treatment of the problem of acceleration of macroscopic projectiles to the velocity regime necessary to obtain the conditions required for thermonuclear fusion ($10^2$–$10^3$ km/s) can be found in the *Proceedings of the Impact Fusion Workshop* held at Los Alamos in 1979.[21] Due to the velocities involved, the only technologies that should be considered are those wherein no contact with the projectile is made, and thus the use of electrostatic or electromagnetic launchers were considered almost exclusively, and this will also be the case in this paper.

## 2.1    Electrostatic Accelerators

Electrostatic accelerators consist of charged particles (either elementary particles or macroscopic pellets) that fall through a voltage potential. The velocity of a particle of mass $m$ accelerated by a voltage potential $\Delta V$ is

$$v = \sqrt{2\frac{q}{m}\Delta V}$$

The particle charge to mass ratio ($q/m$) is the significant parameter that determines the design and capability of different accelerators. Table 1 lists different types of particles that can be accelerated, their charge to mass ratio, and the velocities obtained through a potential of $\Delta V = 1$ MV, which is representative of the maximum voltage that can be realized in a single-stage device. Note that a potential of 1 MV is sufficient to accelerate a proton to nearly 5% of light speed, a velocity that was obtained in the earliest particle accelerators built by Van de Graaff and Cockcroft and Walton. Similarly, any fully ionized atom can be accelerated to similar velocities, since the charge-to-mass ratio is half (or a bit less) than that of a proton due to the presence of neutrons in the nucleus. It is for this reason that the study of the impact of atoms and ions at velocities as great as $10^7$ m/s are encountered in multi-MeV sputtering, as discussed in the Introduction.

The phenomenon ultimately limiting the gradient of the potential field in an electrostatic accelerator is *field emission*, the profuse streaming of electrons from a surface via quantum tunneling through the potential well that has been distorted by the applied voltage gradient. The onset of significant field emission occurs at a gradient on the order of $10^9$ V/m. Field emission also limits the charge-to-mass ratio of the macroscopic object being accelerated, and larger particles are limited in their charge by the strength of the material, as discussed in the Appendix. The maximum velocity attainable via a 10 MV voltage potential is shown as a dashed line in Figure 1 and forms an upper bound on the velocities of macroscopic projectiles obtained with various single stage accelerator technologies. In order to accelerate a μm-sized particle representative of an interstellar dust grain, the grain will have a charge to mass limit of $q/m = 20$ C/kg. If a potential of $10^9$ V/m were applied along the length of the entire accelerator, then in order to accelerate a 1 μm dust grain to 0.1 $c$ (30,000 km/s) would require an accelerator of length 22.5 km, which is impractical.

Cyclotrons and Synchrotrons (e.g., Tevatron, RHIC, and LHC) have accelerated fundamental particles, including the nuclei of heavy (gold and lead) atoms, to very near light speed by successively applying a voltage potential driven by radio frequencies while the nuclei follow a circular track in a static magnetic field. The idea that this approach might be considered to accelerate a μm-sized object by similarly employing a circular accelerator is appealing, since the oscillating voltage potential is essentially reused with each trip around the track. The magnetic field strength is limited by the mechanical strength of the magnets and quenching of the superconductors used to create the magnetic field. A circular accelerator with a magnetic field comparable to the RHIC or LHC (3 to 8 T) would require a radius measuring $r = v/[(q/m) B] \approx 1000$ km to turn μm-sized particles with $q/m \approx 20$ C/kg traveling at $v = 0.2c$, which is likely not feasible on earth. Construction of large particle accelerators in space has been proposed, discussed in Varley et al. [22], and several advantages of the concept are discussed therein, but this is likely as significant an undertaking as interstellar travel itself, if not more so. Further, while particle accelerators are a highly developed technique for accelerating elementary particles with known and fixed charge, it is unlikely that the techniques used can be directly translated over to dust grains for which charge is not known a priori and not likely to remain constant during the acceleration process.

Despite these limitations, it is still of interest to examine current state of the art in electrostatic accelerators, as their capabilities might be of interest for experimental simulation of dust impacts at lower velocities and for smaller sized grains. In recent decades, $C_{60}$ molecules, larger clusters of gold atoms, and biomolecules have been accelerated using the voltage potential generated by electrostatic tandem accelerator facilities.[23, 24] Much of this work has been motivated by mass spectroscopy and secondary ion mass spectrometry of the target material in particular. Molecules and clusters accelerated include $C_{60}^{+2}$ and $Au_{400}^{+4}$, but their charge-to-mass ratios, being only partially ionized compared to elementary charged particles, result in velocities of 3000 km/s and 300 km/s, respectively, for a 10 MeV potential accelerator.

Matrix-assisted laser desorption/ionization (MALDI) is a more recent technique applied to mass spectrometry to characterize large biomolecules (e.g., proteins) that benefits from greater ion velocities for greater detection efficiency. To improve this technique for molecules of greater mass, Hsu et al. [25] have recently developed a *macromolecular ion accelerator* (MIA) that uses voltages applied to plates by fast high voltage switches via a preprogramed waveform generator. This modest device, with a total length of 1 m with an average potential gradient of 1 MV/m, is able to accelerate singly charged immunoglobulins (masses in the 100,000s of atomic mass units or approximately $10^{-22}$ kg) to velocities of 35 km/s, and this technique could be extended up to the limit of high voltage switches.

Moving up the mass scale to dust particles in the 10 nm to 1 um size range, Van de Graaff accelerators have seen application to studying impact of interplanetary dust grains for more than 50 years. The original facility built by Friichtenicht in 1962 was able to accelerate 0.1 μm iron spheres to 14 km/s using a 2 MV potential. This original dust accelerator was the basis for a number of active accelerator facilities operating around the world today, including at the University of Kent (Canterbury, UK) [26], University of Colorado (Boulder, USA) [27], and Max Planck Institut für Kernphysik in (Heidelberg, Germany) [28]. These facilities can continuously accelerate dust grains from micron size down to 30 nm size from velocities of 1 to 100 km/s, respectively, at a rate of about one grain per second. The modern facilities have the ability to actively select particles based on velocity, charge, or mass. The maximum velocities obtained are a combination of the charge on the particle (again, field emission limited) and detection limits (i.e., it may be possible that nanometric grains are being accelerated to greater speeds but cannot be detected). A multistage version of a Friichtenicht-style dust accelerator was proposed by Vedder but did not demonstrate velocities greater than single stage devices. [29]

To conclude this discussion, a few concepts—yet to be demonstrated—are mentioned that might have the potential to overcome the limits outlined here. In order to overcome the limitation on the potential gradient of $10^9$ V/m imposed by field emission, Winterberg [30-32] suggested the concept of magnetic insulation, wherein a current is used to create a local magnetic field that would trap and return electrons emitted by field emission to the cathode surface and thereby insulating against breakdown. Harrison [33] suggested the limit on the charge-to-mass ratio of a projectile might be increased by using a needle-shaped projectile that, due to the concentration of charge at the tip, would experience massive field emission of electrons, resulting in a highly positively charge projectile. In effect, this *field emission projectile* idea turns field emission into an advantage rather than a limit. Liu and Lei [34], drawing upon earlier ideas originating from Harrison [35], proposed blowing a continuous stream of ions onto a projectile, generating a local high-strength electric field. None of these concepts—all of which were motivated by fusion research—have been experimentally demonstrated to generate greater projectile velocities.

## 2.2 Electromagnetic Accelerators

The limitations discussed, which suggest a launcher capable of accelerating 1 g to 1000 km/s would be tens of km long, motivated the Los Alamos 1979 Impact Fusion Workshop to consider predominately electromagnetic launchers that utilize the Lorentz force ($j \times B$) to accelerate the projectile. The simplest implementation is the railgun, wherein the projectile comprises the armature for current flowing through the rails that make up the side walls of the launcher. Current flowing up one rail and down the other create the driving magnetic field that exerts a Lorentz force on the current flowing through the projectile. Despite intensive work on rail guns since the 1970s, the maximum velocity of a projectile retaining integrity obtained with a railgun is about 6 km/s, less than conventional light gas guns for the equivalent projectile mass. The use of sliding contact between the projectile and rails would preclude their application for the velocities of interest for this article. Contactless accelerators, such as coilguns, would energize and/or de-energize coils in synchronization with the projectiles, itself a solenoid (either inductive or superconducting). There are a number of embodiments this concept can take, but to date the maximum velocities demonstrated by coilguns are less than 1 km/s. The reasons for these limited velocities are related to the complexities of switching on and off high currents and beyond the scope of this article. The interested reader can find the engineering details of the electromagnetic launchers in the proceeding of the International Symposia on Electromagnetic Launch Technology, most of which are published in the January issue of *IEEE Transactions on Magnetics* and *IEEE Transactions on Plasma Science* in odd-numbered years.

## 2.3 Explosive Accelerators

The generation of hypervelocity jets of metal from metal-lined explosive cavities, i.e., a *shaped charge*, is a well-developed military technology that also sees civilian application, for example, in perforating the casing in gas and oil drilling operations. The mechanism by which these devices operate was explained in a classic paper by Birkhoff, Taylor, and co-workers [36] and extended in a model by Pugh et al. [37] Due to the high pressures that greatly exceed the yield strength of the metal, the collapse of the metal liner can essentially be treated as a flow of liquid toward the centerline that, by conservation of mass and momentum, requires some of the liner be jetted forward. This mechanism is essentially the same as that which results in the generation of a jet when a cavity in water collapses. An interesting feature of the model is that, as the angle of the cavity is made smaller and the phase velocity at which the implosion of the cavity is increased, the velocity of the jet produced can be made arbitrarily great. This motivated Koski et al. [38] to examine the collapse of evacuated metal tubes by imploding toroidal detonations onto them. Jet velocities as great as 100 km/s were observed via streak photography, although the nature of these jets was not determined. Later, Lunc et al. [39] proposed to use a two-stage jet formation process in which an imploding toroidal detonation imploded a tube onto a jet originating from an earlier implosion of the same tube. Again, velocities of 100 km/s were reported. Although the nature of these jets is a mystery, the analysis outlined in Koski and Lunc et al., would suggest these metal jets are likely of the order of $10^{-3}$ g, making them the most massive projectiles that have been launch to date by technological means to 100 km/s. These devices are unlikely to be able to be extended beyond 100 km/s and the fact that the projectile is not well quantified makes them less than ideal for impact testing. Explosive accelerators that retain an intact projectile exist and are reviewed in [40], but are likely limited to velocities less than 20 km/s. The low cost and the demonstrated ease of scaling shaped charges (scaling as jet mass ~ explosive mass over several orders of magnitude) might still make them attractive for examining impacts of massive projectiles at these velocities.

## 3. High Energy Density Facilities

Drivers used for experiments in *high energy density physics* (not to be confused with high-energy physics) are motivated by interest in inertial confinement fusion, nuclear weapon simulation, fundamental equation of state studies, and experimental astrophysics. These generally fall in two classes: pulsed power and pulsed lasers. These facilities are also capable of accelerating macroscopic objects to velocities approaching, and in some cases exceeding, 100 km/s in very short distances (usually, mm or less). Due to the enormous driving pressures generated, the projectile cannot be allowed to undergo free flight (like the projectiles accelerated by the techniques covered in Section 2) because the projectiles would explosively disassemble or vaporize upon release of the driving pressure. Nonetheless, these facilities are among the most promising for the creation of highly energetic events that are relevant to the simulation of the dust grain impact problem for interstellar flight.

## 3.1 Pulsed Power

The most powerful pulsed power facility in existence is the Z-Machine at Sandia National Laboratory, with capacitor banks that store up to 20 MJ of energy that can be released on 100-600 ns timescales, generating currents of 25 MA.[41] The magnetic field generated (1200 T) results in magnetic pressures ($B^2/2\mu_0$) as great as 600 GPa. When applied to thin metallic foils, the magnetic pressure can accelerate the foils (typically ≈ 2 cm × 1 cm × 1 mm) to velocities approaching 50 km/s over travel distances of less than 1 cm. As the magnetic field diffuses into the foil being accelerated, it is heated be Joule heating and vaporizes. In addition, the ramp-up of the magnetic field must be sufficiently slow to avoid the formation of a shock within material being accelerated. These two effects set an ultimate velocity limit on the order of 100 km/s that can be achieved via the application of magnetic pressure.[42] The impact of the metal foil flyers launched by the Z-Machine at velocities of 27 km/s has been used, for example, to shock samples of water to the regime of pressure and temperature encountered at the center of ice giant planets.[43]

### 3.2 Pulsed Lasers

The use of very high power pulsed lasers for internal confinement fusion (ICF) has motivated application of these devices to launching thin flyers to velocities exceeding 100 km/s. In ICF, laser ablation of the exterior of a condensed-phase deuterium-tritium pellet (mm-scale) drives a compression pulse or shock into the pellet at speeds on the order of 300 km/s, in a manner similar to rocket propulsion, raising the density by a factor of a thousand, in the hope of achieving fusion conditions. In recent decades, consideration of *fast ignition* concepts that would use an additional energetic event (e.g., second laser pulse) to ignite the compressed pellet has received significant attention. The use of flyers launched by the laser as the fast ignition source when impacting upon the previously compressed fuel is one such fast ignition concept (note this is separate from impact fusion discussed in the Introduction). [44, 45] Experiments using the GEKKO XII laser in Japan [44] and the Nike krypton fluoride laser at the Naval Research Laboratory in the US [45] have used laser-driven ablation of polystyrene foils (0.5 mm diameter, 10 to 20 μm thick) to velocities of 700 and 900 km/s, respectively. These studies estimated that approximately 90% of the foil mass was lost in ablation, and the density and state of the remaining material was difficult to estimate. Nonetheless, as seen in Figure 1, these tests represent the nearest, in terms of mass and velocity, experiments to the more significant threat represented by the impact of μm-sized dust grains.

As discussed by Badziak et al. [46], use of laser-driven plasma ablation has poor efficiency in these experiments and the technique is unlikely to be able to extend launch of foils above 1000 km/s, however, using radiation pressure generated in the laser-driven plasma might allow greater velocities to be achieved. This idea is discussed further by Macchi et al.[47-49], essentially a laser light sail in the lab driven by the radiation created by laser energy deposition. Interestingly, the most recent discussion of this concept mentions it in connection with Breakthrough Starshot, since both concepts rely on radiation pressure to approach relativistic velocities.[49]

### 4. Discussion

As accelerators do not appear to be able to achieve the regime of projectile mass and velocity necessary to perform laboratory simulation of dust grain impacts at starflight velocities, it is of interest to explore other types of facilities that could be used to create conditions similar to those encountered in such impacts. In particular, if the correct energy and energy release rate (power) can be realized, then such facilities might be useful for the study of the consequences of such impact on spacecraft structures.

The more significant challenge becomes the power density requirement, i.e., the rate of energy density. The energy density deposition time can be estimated for a 1 μm dust grain as the transit time across the grain, which at $0.1c$ is $10^{-13}$, or 0.1 ps (100 fs), which is accessible via mode-locking techniques for pulsed lasers and, for even greater energies, via chirped pulse amplification. Thus, rather than use these high-power pulsed laser facilities to launch a projectile (as discussed in Section 4), they could be used to create intense events with the correct power and energy densities as a grain impact, and then representative coupons of spacecraft materials (e.g., sail, etc.) located at various distances could be instrumented to see their time-resolved response. By examining the response seen at different time intervals and different distances, the structural response that would be obtained if the impact fireball was receding at speeds of $0.1c$ (or less) could be reconstructed.

## 5. Conclusions

The interstellar medium encountered at velocities necessary for interstellar travel will impact the spacecraft at energies on the order of MeV per nucleon. For elementary particles, due to their large charge-to-mass ratios (i.e., e/Da ~ 1), conventional accelerators operating at MV potentials have been able to reproduce impacts at these energies in the lab for over a century, and thus a wealth of experimental data already exists to address sputtering by interstellar atom and ions. For molecules, atomic clusters, and progressively larger dust grains, the velocity that objects can be accelerated to in the laboratory via electrostatic techniques becomes progressively less as the mass increases, due to the decreasing charge to mass ratio. Charge is limited by field emission and the ultimate strength of the materials involved, and this fundamental limit is not easily overcome; as a result, nanometric to micron-sized dust grains can presently only be accelerated to 100 km/s in the laboratory, or 0.03% lightspeed, but still represent the fastest well-quantified macroscopic objects on Earth. Contactless electromagnetic launchers such as coilguns, utilizing the Lorentz force, have not exceeded 1 km/s. Other technologies, such as explosive accelerators and pulse-power facilities, may be able to contribute by accelerating larger, but poorly quantified, masses (perhaps up to a gram) to velocities exceeding 100 km/s, but likely not greater. Presently, the greatest velocity that can be obtained at near-condensed matter densities is via pulsed laser facilities predominately used for inertial confinement fusion research. Using laser-driven ablation, velocities of almost 1000 km/s (0.3% light speed) have been obtained, although the state of the matter accelerated cannot be precisely quantified and the projectiles cannot be allowed to travel over distances that would permit clean impact experiments to be conducted, independent of the laser energy deposition. Even greater velocities might be attained using radiation-pressure driven acceleration, essentially, laser-driven light sails in the lab, a concept that is presently being explored.

Since the prospect for obtaining velocities on the order of 10,000 km/s in the lab is doubtful, it is proposed instead to use pulsed laser facilities to create dense plasmas that can mimic the power and energy densities of impact events without launching actual projectiles. The damage caused by these events on spacecraft and sail can then be studied at scaled distances from the event. By integrating damage accrued over specific time intervals, the motion of the plasma source moving away from the target at high velocity can be simulated in these experiments. A test following this procedure would represent the best terrestrial-based simulation of the impact events likely to be encountered during interstellar flight.

**Appendix: Limit on charge to mass ratio of a macroscopic object**

**Field Emission**
We consider a charged, conductive sphere of radius $r$ and surface charge $\sigma$. The total charge is thus $q = 4\pi r^2 \sigma$. The potential outside a sphere is $V = \frac{q}{4\pi\varepsilon_0 r}$, resulting in a potential gradient at the surface of the sphere is
$$\frac{dV}{dr} = -\frac{q}{4\pi\varepsilon_0 r^2}$$
If the limit of gradient at which the onset of significant field emission is $\left(\frac{dV}{dr}\right)_{FE} = 10^9$ V/m, then the maximum charge to mass ratio for a solid sphere can be expressed as
$$\left(\frac{q}{m}\right)_{sphere} = \frac{3\varepsilon_0}{r\rho}\left(\frac{dV}{dr}\right)_{FE}$$
If the sphere is a shell with wall thickness to radius ratio $\delta = t/r$, then the charge to mass ratio is
$$\left(\frac{q}{m}\right)_{shell} = \frac{\varepsilon_0}{t\rho}\left(\frac{dV}{dr}\right)_{FE} = \frac{\varepsilon_0}{r\,\delta\rho}\left(\frac{dV}{dr}\right)_{FE}$$
Note that the limiting charge-to-mass ratio of a shell of fixed thickness $t$ is independent of sphere radius. These relations are plotted in Figure 2 for values of $\rho = 2200$ kg/m$^3$, representative of an advanced carbon or aluminum/magnesium/lithium material.

**Material Strength**

The other factor limiting charge-to-mass ratio is the strength of the particle material. If the surface of the thin-walled spherical shell has charge $q$, the tensile stress in the shell due to electrostatic repulsion is

$$\sigma_t = \frac{q^2}{32\, t\, r^3 \pi^2 \varepsilon_0}$$

For advanced engineering materials, $\sigma_{\text{ult}} \approx 10^9$ Pa = 1 GPa. Thus, the charge-to-mass ratio limited by material strength is given by

$$\left(\frac{q}{m}\right)_{\text{strength}} = \sqrt{\frac{2\varepsilon_0 \sigma_{\text{ult}}}{\rho^2 r\, t}} = \frac{1}{\rho r}\sqrt{\frac{2\varepsilon_0 \sigma_{\text{ult}}}{\delta}}$$

This relation is also plotted in Fig. 2. For particles of micron size and smaller, field emission sets the limit of the maximum charge-to-mass ratio that can be achieved. Depending on the shell thickness, material strength limits the charge that can be held on a larger spherical shell. The limiting value of $q/m$ determines the velocity that can be obtained in an electrostatic accelerator and the radius through which a particle can be turned in a magnetic field. A more detailed analysis of these considerations can be found in [50].

Table 1 Properties of particles that can be accelerated by an electrostatic field

| Particle | Mass (kg) | Charge (C) | $q/m$ (C/kg) | $q/m$ (e/Da) | Velocity via 1 MV Potential $v$ (m/s) | Velocity via 1 MV Potential $v/c$ | Gyroradius $v/c = 0.1$, $B = 1$ T |
|---|---|---|---|---|---|---|---|
| Proton | $1.67 \times 10^{-27}$ | $1.60 \times 10^{-19}$ | $9.58 \times 10^{7}$ | 1.0 | $1.38 \times 10^{7}$ | 0.046 | 0.31 m |
| Au nucleus | $3.27 \times 10^{-25}$ | $1.27 \times 10^{-19}$ | $3.87 \times 10^{7}$ | 0.4 | $8.80 \times 10^{6}$ | 0.029 | 0.77 m |
| $C60^{+2}$ | $1.20 \times 10^{-24}$ | $3.20 \times 10^{-19}$ | $2.68 \times 10^{5}$ | 0.0027 | $7.32 \times 10^{5}$ | 0.0024 | 1.1 m |
| $Au_{400}^{+4}$ | $1.31 \times 10^{-22}$ | $6.41 \times 10^{-19}$ | $4.90 \times 10^{3}$ | $5 \times 10^{-5}$ | $9.90 \times 10^{4}$ | $3.3 \times 10^{-4}$ | 6.1 km |
| Biomolecules (immunoglobulins, etc.) | $2.5 \times 10^{-22}$ $6.4 \times 10^{-22}$ | $1.60 \times 10^{-19}$ | 640 250 | $2.6 \times 10^{-6}$ $2.6 \times 10^{-6}$ | $36 \times 10^{3}$ $22 \times 10^{3}$ | $1 \times 10^{-4}$ $7 \times 10^{-5}$ | 47 km 120 km |
| Dust grain ($\rho = 2.5$ g/cm$^3$) | | | | | | | |
| $d = 10$ nm | $1.3 \times 10^{-21}$ | $2.7 \times 10^{-18}$ | 2000 | $2 \times 10^{-5}$ | $65 \times 10^{3}$ | $2 \times 10^{-4}$ | 15 km |
| $d = 1$ μm | $1.3 \times 10^{-15}$ | $2.7 \times 10^{-14}$ | 20 | $2 \times 10^{-7}$ | $6.5 \times 10^{3}$ | $2 \times 10^{-5}$ | 1500 km |
| $d = 100$ μm | $1.3 \times 10^{-9}$ | $2.7 \times 10^{-10}$ | 0.2 | $2 \times 10^{-9}$ | $0.65 \times 10^{3}$ | $2 \times 10^{-6}$ | 150,000 km |

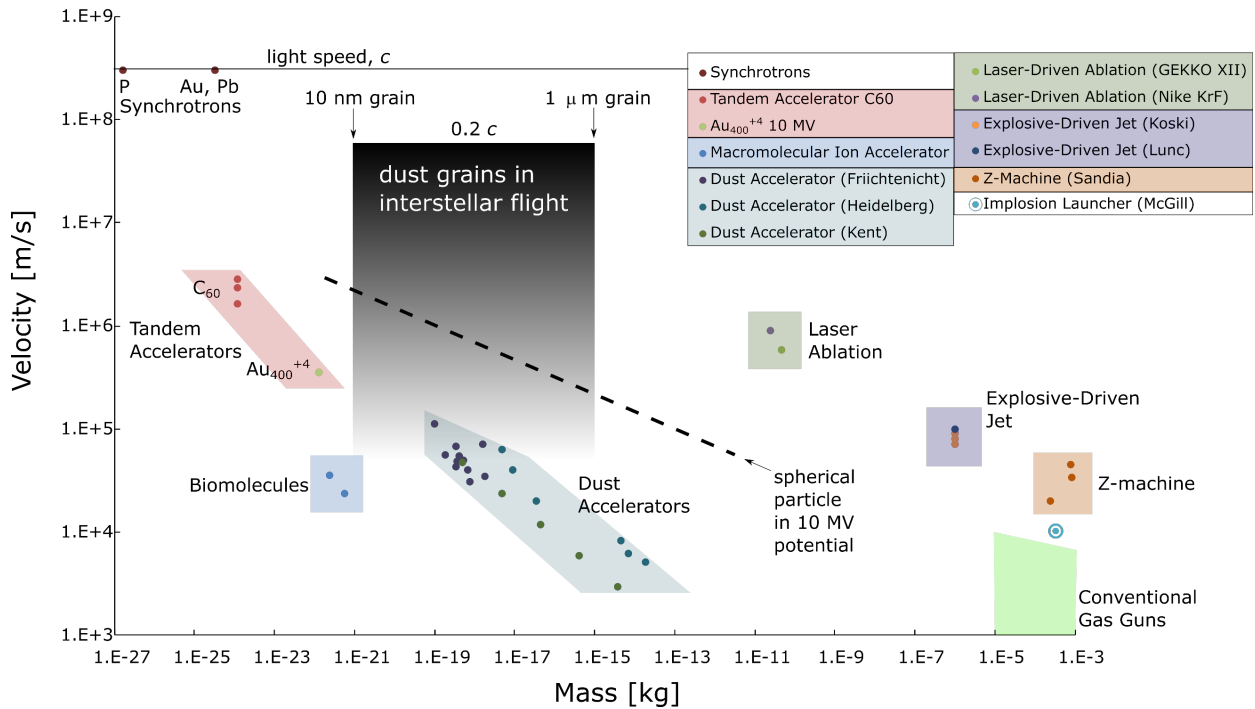

Figure 1: Overview of accelerator technologies and their capability in terms of particle mass and velocity. The dashed curve represents the velocity of a spherical conductive particle with $\rho = 2500$ kg/m$^3$ with charge-to-mass ratio limited by field emission at the surface critical gradient of $10^9$ V/m and accelerated through a 10 MV potential.

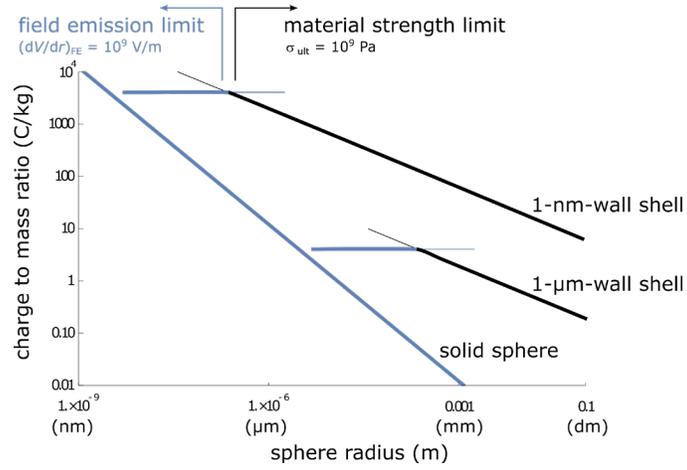

Figure 2: Charge to mass ratio (C/kg) for spherical particles as a function of size. Field emission due to the voltage gradient at the particle surface sets a limit for smaller particles. Larger particles made of thin-walled shells are limited by material strength.